\documentclass[aps,prl,twocolumn,superscriptaddress,showpacs,amsmath,amssymb,preprintnumbers]{revtex4}

\usepackage{graphicx,amssymb}
\begin{document}


\title{Spin Configuration in the 1/3 Magnetization Plateau of Azurite Determined by NMR}

\author{F. Aimo}
\affiliation{Grenoble High Magnetic Field Laboratory, CNRS, BP 166, 38042
Grenoble Cedex 9, France}

\author{S. Kr\"{a}mer}
\affiliation{Grenoble High Magnetic Field Laboratory, CNRS, BP 166, 38042
Grenoble Cedex 9, France}

\author{M. Klanj\v{s}ek}
\altaffiliation[Present address: ]{Jo\v{z}ef Stefan Institute, Jamova cesta 39,
1000 Ljubljana, Slovenia} \affiliation{Grenoble High Magnetic Field Laboratory,
CNRS, BP 166, 38042 Grenoble Cedex 9, France}

\author{M. Horvati\'{c}}
\altaffiliation[Email address: ]{mladen.horvatic@grenoble.cnrs.fr}
\affiliation{Grenoble High Magnetic Field Laboratory, CNRS, BP 166, 38042
Grenoble Cedex 9, France}

\author{C. Berthier}
\affiliation{Grenoble High Magnetic Field Laboratory, CNRS, BP 166, 38042
Grenoble Cedex 9, France}

\author{H. Kikuchi}
\affiliation{Department of Applied Physics, University of Fukui, Fukui
910-8507, Japan}

\date{\today}

\begin{abstract}
High magnetic field $^{63,65}$Cu NMR spectra were used to determine the local
spin polarization in the 1/3 magnetization plateau of azurite,
Cu$_3$(CO$_3$)$_2$(OH)$_2$, which is a model system for the distorted diamond
antiferromagnetic spin-1/2 chain. The spin part of the hyperfine field of the
Cu2 (dimer) sites is found to be field independent, negative and strongly
anisotropic, corresponding to $\approx$10\,\% of fully polarized spin in a
$d$-orbital. This is close to the expected configuration of the ``quantum''
plateau, where a singlet state is stabilized on the dimer. However, the
observed non-zero spin polarization points to some triplet admixture, induced
by strong asymmetry of the diamond bonds $J_1$ and $J_3$.
\end{abstract}



\maketitle

\begin{figure}[b]
\includegraphics[width=1.00\linewidth,clip]{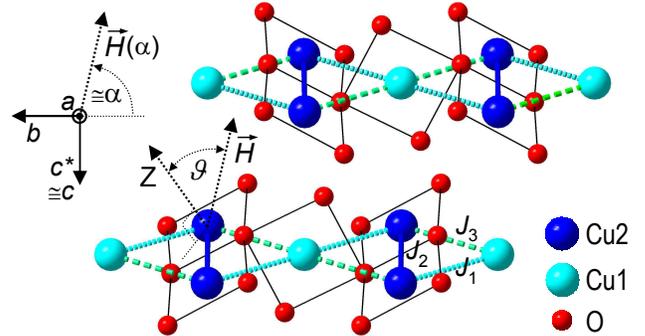}
\caption{\label{fig:one} (color online) Diamond chains formed by the exchange
interactions, $J_{ij} \mathbf{S}_i\cdot\mathbf{S}_j$, between $S=1/2$ spins of
Cu$^{2+}$ ions in the crystal structure of azurite. There are two equivalent
but differently oriented chains, related by the $ac$-plane of mirror symmetry.
Chains contain ``monomer'' spins on the Cu1 sites 
coupled by $J_1$ 
and $J_3$ 
interactions to each spin of the dimer formed by the two Cu2 sites
, mutually coupled by $J_2$
. There is an inversion symmetry on each Cu1 site and at the center of each
dimer. For each
Cu site 4 nearest neighboring oxygen atoms (
connected by thin lines) define approximately the plane of the local symmetry
of the wave functions and of the corresponding EFG tensor. Dotted line vectors
define the rotation angle $\alpha$, and the angle $\vartheta$ between the
magnetic field and the Z principal axis of the EFG tensor. C and H atoms are
not shown.}
\end{figure}

The natural mineral azurite, Cu$_3$(CO$_3$)$_2$(OH)$_2$, has been recently
recognized \cite{Kikuchi_PRL05} as a model system for the frustrated
antiferromagnetic Heisenberg spin-1/2 chain of ``distorted diamond'' geometry
defined in Fig.~\ref{fig:one}. Its most prominent feature is a large plateau in
the magnetization curve at 1/3 of the saturation magnetization, which extends
from 11 to 30\,T when the applied magnetic field ($H_0$) is perpendicular to
the chains. Such a ``1/3 plateau'' is usually associated with a classical
collinear up-up-down ($uud$) type of spin arrangement, or rather to a quantum
state which has this classical analogue. For example, a $uud$ state is
predicted for spins on a two-dimensional triangular lattice and observed in the
Cs$_2$CuBr$_4$ compound \cite{uud}. The 1/3 plateau in azurite is proposed to
be of fundamentally different, ``$00u$'' type, where the dominant $J_2$
coupling ensures that the two ``dimer'' spins on the Cu2 sites (see
Fig.~\ref{fig:one}) are in a singlet state, while the third ``monomer'' (Cu1)
spin is completely polarized by the field. As this state is based on the
presence of a singlet, it is of pure quantum nature without a classical
analogue. Azurite is a good candidate to be the first system exhibiting such a
1/3 plateau state, but a direct experimental evidence is still missing. The
point is that both types of plateaus are predicted for a diamond chain, the
$00u$ type driven by dominant $J_2$ coupling and an analogue of the $uud$ state
in presence of dominant $J_1$ and $J_3$ \cite{Okamoto_JPC03, Gu_PRB07}. The
azurite is close to the phase boundary between them, and there is a controversy
on the $J$ values proposed from the magnetization, specific heat and neutron
scattering experiments \cite{Kikuchi_PRL05, Comment,Rule_PRL08,Mikeska_PRB08}.
The two different plateau types are distinguished by very different local spin
polarizations, which can in principle be directly accessed by performing
nuclear magnetic resonance (NMR) on the on-site copper $^{63,65}$Cu nuclei. In
this letter we present such NMR data which show that in the 1/3 plateau the
dimer spins are nearly in the singlet configuration and thus confirm the $00u$
type of plateau. We find a small non-zero spin polarization of these sites,
estimated to approximately 10\,\% of full polarization, which points to an
important asymmetry of $J_1$ and $J_3$ couplings. The observed polarization is
incompatible with a $uud$ type of plateau, in which the dimer spins are
strongly polarized.

\begin{figure}[t]
\includegraphics[width=1.00\linewidth,clip]{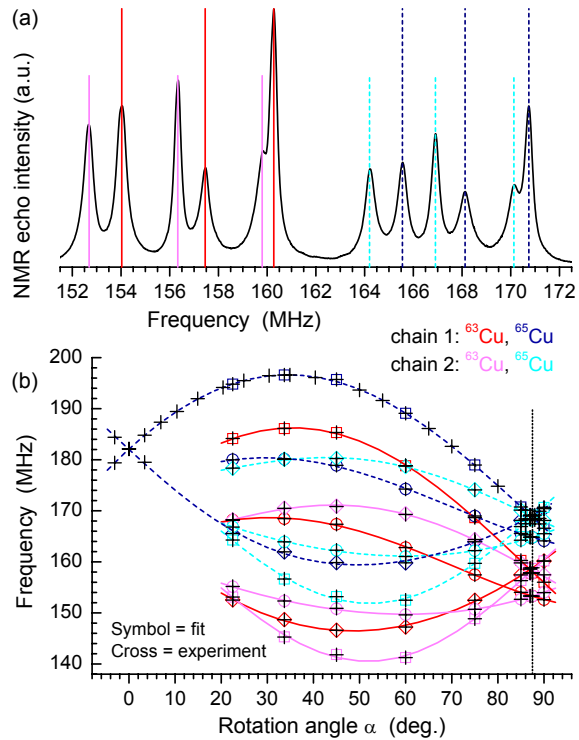}
\caption{\label{fig:two} (color online) Rotation dependence of copper NMR
spectra of azurite at $T$\,=\,1.5\,K. Magnetic field of 15.0\,T is applied
perpendicular to the rotation axis that was close to the crystal $a$-axis. (a)
NMR spectrum taken at $\alpha$\,=\,85$^{\circ}$ and the corresponding fit
(vertical lines), as explained in the text. (b) Angular dependence of the
observed line positions (crosses, with lines to guide the eye) and the
corresponding fits (open symbols: circles for the central transitions and
squares and diamonds for the satellites). Color code is given in the figure:
dark (light) grey lines for the chain 1 (2) and solid (dashed) lines for the
$^{63(65)}$Cu isotope. Vertical dotted line denotes $H_0~||~ac$-plane
orientation, where both chains are identical.}
\end{figure}

In general, the copper NMR spectrum of a single crystal consists of 6 NMR lines
per each non-equivalent Cu site, corresponding to three transitions between
energy levels of a spin $I = 3/2$ nucleus for each of the two $^{63}$Cu and
$^{65}$Cu isotopes. In the crystallographic structure of azurite, shown in
Fig.~\ref{fig:one}, we recognize two different copper sites in two equivalent
chains of different orientation with respect to the arbitrary direction of the
applied magnetic field $\bf{H}_{0}(\alpha)$. We also note that the two Cu2
sites of each dimer are expected to be undistinguishable by NMR, which is
ensured by the inversion symmetry with respect to the center of each dimer. We
thus expect that $^{63,65}$Cu NMR spectrum has
6\,$\times$\,2\,$\times$\,2\,=\,24 NMR lines. Single crystal spectra presented
in Fig.~\ref{fig:two} contain only 12 lines, meaning that we observe only one
of the two Cu sites. We recall that these spectra are taken in the plateau
phase, so that they are not affected by the N\'{e}el ordering appearing at
1.9\,K at field values \textit{below} the plateau phase \cite{Kikuchi_PTPS05}.
A standard way for the identification of the observed site is to compare the
symmetry of the local electric field gradient (EFG) tensor determined by NMR to
what is expected from the local symmetry of the four nearest neighboring (NN)
oxygen atoms (see Fig.~\ref{fig:one}). This rather technical procedure,
explained in detail in the following paragraph, unambiguously demonstrates that
only Cu2, i.e. the dimer site is observed by NMR.

\begin{figure}[t]
\includegraphics[width=1.0\linewidth,clip]{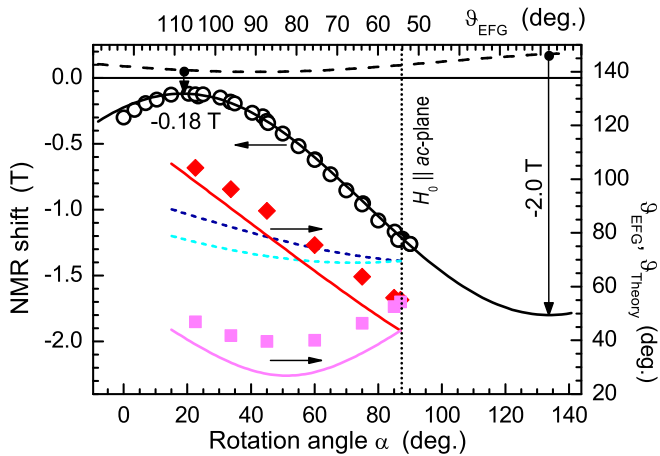}
\caption{\label{fig:three} (color online) Left scale (black lines and symbols):
rotation dependence of the NMR shift obtained from the highest frequency
$^{65}$Cu NMR line (open circles), the fit to these data (solid line) and an
estimate of the orbital shift (dashed line). These results provide an estimate
of the hyperfine field $H_{\rm{spin}}^{\parallel}\approx -2.0$\,T and its
anisotropy $H_{\rm{spin}}^{\parallel} / H_{\rm{spin}}^{\perp}\approx 11$, which
corresponds to 10\,\% spin polarization in a $d$-orbital. Right scale
(color\,/\,grey-scale lines and symbols): $\vartheta_{EFG}$ deduced by fitting
the line positions shown in Fig.~\ref{fig:two}(b) (solid squares and diamonds),
following closely the predictions for the dimer Cu2 site (solid lines) and not
the
Cu1 site (short-dashed lines), as explained in the text. Experimental
$\vartheta_{EFG}(\alpha)$ dependence (diamonds) is used to define the upper
horizontal scale.}
\end{figure}

Each copper isotope ($I = 3/2$) generates a triplet of NMR lines whose average
frequency reflects the Zeeman coupling to the total ``effective'' magnetic
field ${\bf{H}}_{\rm{eff}}$, while the line splitting is induced by the
``quadrupolar'' coupling to the local EFG. The corresponding nuclear spin
Hamiltonian, ${\cal{H}} = \hbar \gamma {\bf{I}} \cdot {\bf{H}}_{\rm{eff}} + h
\nu_{\rm{Q}} [3I_{\rm{Z}}^2 -I(I+1)+ \eta (I_+^2 + I_-^2)/2]/6$, is uniquely
defined by 5 parameters: the EFG tensor described by the quadrupolar coupling
$\nu_{\rm{Q}}$ and its asymmetry parameter $\eta$, and $H_{\rm{eff}}$ and its
direction ($\vartheta_{\rm{EFG}}$, $\varphi_{\rm{EFG}}$) with respect to the
principal axes (X,Y,Z) of the EFG tensor \cite{Abragam61}. Knowing the
gyromagnetic ratios for the two isotopes, $^{63}\gamma$ and $^{65}\gamma$, as
well as the ratio of their quadrupolar couplings,
$^{63}\nu_{\rm{Q}}/^{65}\nu_{\rm{Q}}$\,=\,1.0805, these 5 parameters can be fit
to provide the observed 6 NMR frequencies (for each chain). The $\nu_{\rm{Q}}$
and $\eta$ parameters do not depend on the orientation of magnetic field and,
in particular, they are common to spectra from two chains shown in
Fig.~\ref{fig:one}. In this experiment we performed in-situ rotation of the
crystal around the axis that was close to the crystal $a$-axis, and have taken
several complete spectra at different rotation angles $\alpha$ (see
Fig.~\ref{fig:one}), as shown in Fig.~\ref{fig:two}. The smallest NMR line
widths and thus the most precise fits are obtained when $\vartheta_{\rm{EFG}}
\cong 90^{\circ}$, which for chain 1 corresponds to $\alpha \cong 45^{\circ}$.
This particular orientation has therefore been used to determine
$^{63}\nu_{\rm{Q}}$\,=\,36.5\,MHz and $\eta$\,=\,0.085, and the obtained EFG
values have been successfully used to produce \textit{all} the other fits -
this time by fitting only three parameters ($H_{\rm{eff}},
\vartheta_{\rm{EFG}}$, $\varphi_{\rm{EFG}}$) for each set of 6 NMR frequencies
(see Fig.~\ref{fig:two}). The experimentally obtained
$\vartheta_{\rm{EFG}}$$(\alpha)$ dependence could then be compared to a simple
approximate estimate for this quantity, based on the crystallographic
structure. We know that for an ideal tetragonal coordination a pure
$d_{{\rm{X}}^2-{\rm{Y}}^2}$ orbital pointing towards 4 NN oxygens generates
axially symmetric EFG with the strongest principal axis along the Z direction.
Therefore, the best estimate for the Z axis is the normal to the plane
approximately defined by 4 NN oxygens (see Fig.~\ref{fig:one}). The direction
of ${\bf{H}}_{\rm{eff}}$ is approximated by the direction of the applied field,
supposing that the rotation axis is precisely the $a$-axis of the crystal.
These estimates of Z and ${\bf H}_{\rm eff}$ directions define the angle
$\vartheta_{\rm{Theory}}$, and its rotation  dependence
$\vartheta_{\rm{Theory}}(\alpha)$ is plotted in Fig.~\ref{fig:three} for both
Cu sites and both chains, in comparison with the experimental
$\vartheta_{\rm{EFG}}(\alpha)$ values. Neglecting small offset due to various
approximations, from these data one clearly identifies that the observed NMR
signal corresponds to dimer Cu2 sites and is incompatible with the Cu1 sites.

The principal information obtained from the fits is the spin part of the
hyperfine field, ${\bf{H}}_{\rm{spin}} = {\bf{H}}_{\rm{eff}} -
(1+{\sf{K}}_{\rm{orb}})\,{\bf{H}}_0 = {\sf{A}}\,g \mu_{\rm{B}} \langle {\bf{S}}
\rangle$, induced by the local spin polarization $\langle {\bf{S}} \rangle$
through the hyperfine coupling tensor ${\sf{A}}$. The orbital (Van Vleck) shift
tensor ${\sf{K}}_{\rm{orb}}$ is here a minor correction, because typical
${\sf{A}}$ values for a copper spin are as large as $A_{\rm{Z}}\,\approx
-$20\,T/$\mu_{\rm{B}}$, with large anisotropy $A_{\rm{Z}} / A_{\perp} \approx
10$. Knowing that the EFG and the hyperfine shift tensors are dominantly
determined by the same wave function, we expect that the principal axes of both
tensors are approximately the same, so that the $H_{\rm{spin}}$ vs.
$\vartheta_{\rm{EFG}}$ dependence provides a complete information on the local
spin polarization. In Fig.~\ref{fig:three} we have plotted the rotational
dependence of the experimental NMR line shift ($H_{\rm{eff}}$\,$-$\,$H_0$) and
its extrapolation by a sinusoidal fit, together with an estimate of the orbital
shift ($K_{\rm{orb}}^{\rm{Z}}\approx$ 1.3\,\%, $K_{\rm{orb}}^{\perp}\approx$
0.3\,\%). From these data we estimate the spin-induced hyperfine field to be
$H_{\rm{spin}}^{\parallel}\approx -2.0$\,T, with an anisotropy
$H_{\rm{spin}}^{\parallel} / H_{\rm{spin}}^{\perp}\approx 11$ (where
$\parallel$ and $\perp$ refer to the principal axes of this tensor). This
corresponds to about 10\,\% spin polarization of a typical Cu$^{2+}$
$d_{{\rm{X}}^2-{\rm{Y}}^2}$ orbital. The error in these values is estimated to
be $\approx$\,20\,\%, dominantly from the extrapolation of the angular
dependence to $H_{\rm{spin}}^{\parallel}$ \cite{approx}. In particular, in
Fig.~\ref{fig:three} we clearly see that the maximum of the experimental line
shift is shifted by 18$^{\circ}$ from the expected $\vartheta_{\rm{EFG}} =
90^{\circ}$ value. This means that the principal axes of the EFG and the
hyperfine tensors are not really parallel, pointing to a departure from the
simplified picture of pure $d_{{\rm{X}}^2-{\rm{Y}}^2}$ orbital. Indeed, the
electronic density observed by x-ray diffraction suggest significant admixture
of other orbitals \cite{Belokoneva_01}.

\begin{figure}[b]
\includegraphics[width=1.0\linewidth,clip]{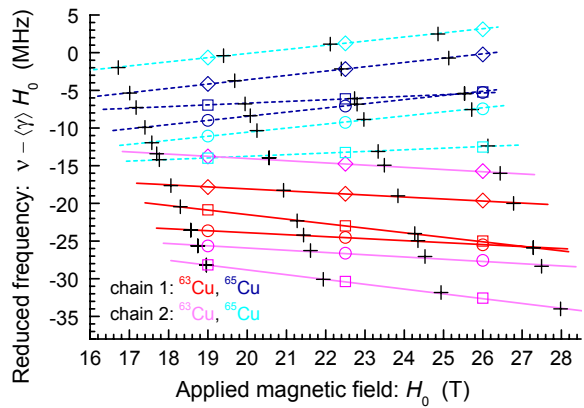}
\caption{\label{fig:four} (color online) Magnetic field dependence of the
$^{63,65}$Cu NMR line positions in azurite at 1.4\,K, for the field orientation
close to the $c$-axis. The spectra were taken at constant frequency by sweeping
the field, and results plotted in the reduced frequency scale to eliminate the
dominant field dependence $\langle\gamma\rangle H_0$, with $\langle \gamma
\rangle$~= ($^{63}\gamma$\,+\,$^{65}\gamma$)/2 = 11.687\,MHz/T. Lines are
linear fits to the observed field dependence, and symbols fits to this linear
interpolation at 19, 22.5 and 26~T, which confirm that the spin polarization of
the dimer Cu2 site is magnetic field independent. Color/symbol code is the same
as in Fig.~\ref{fig:two}.}
\end{figure}

In a true magnetization plateau the magnetization should not vary with the
magnetic field. In order to test this most prominent feature of the plateau in
azurite, we performed very high field measurements of the copper NMR spectra,
in the field range 17--28\,T and at 1.4\,K, for $H_0$ applied close to the
$c$-axis. Because the effects of the quadrupolar coupling and the hyperfine
shift are entangled in the NMR spectra, the shift can only be determined by the
complete NMR fits as explained in the third paragraph. The line positions and
fits shown in Fig.~\ref{fig:four} indeed confirm that $H_{\rm{spin}}$ is to a
high precision field independent in the plateau. For a field variation from 19
to 26\,T (i.e. 37\,\%) the change in the measured $|H_{\rm{spin}}|$ is found to
be (1\,$\pm$\,1)\,\%, where the precision is limited by our estimate of the
orbital shift tensor. This information might be important to constrain the
possible effects of Dzyaloshinski-Moria (DM) interaction terms, which may
induce some weak field dependence of the spin polarizations. The DM interaction
on the dimer bond has been invoked to explain strong anisotropy of the width of
the plateau \cite{Kikuchi_PRL05}. However, the presence of an inversion center
at the center of the dimer precludes such a term, and only DM interaction on
$J_1$ and $J_3$ exchange paths are possible. Whether or not they can explain
the observed anisotropy has not yet been studied theoretically.

Despite considerable efforts, we could not observe the NMR signal from the
monomer Cu1 spin. We have tried to find it at very low temperature in order to
minimize the spin fluctuations. In this way the longitudinal ($T_1^{-1}$)
relaxation rate was reduced, however, the same does not necessarily apply to
the transverse ($T_2^{-1}$) relaxation \cite{fast_T_2}. To provide the correct
total polarization of the 1/3 plateau, the spin polarization at the Cu1 site
has to be $\approx$\,80\,\%, that is $\approx$8 times more than at the dimer
sites. We can then roughly estimate that the corresponding $T_2^{-1}$ ratio is
of the order of 8$^2$\,=\,64. As the dimer $T_2$ values are in the
10--100\,$\mu$s range (depending on the orientation), this factor is enough to
reduce the Cu1 $T_2$ below the experimental dead time for the observation of an
NMR signal. This provides a reasonable explanation why Cu1 spin could not be
observed, but also an important hint on the system: the longitudinal spin
fluctuations (effective in $T_2$ relaxation) are probably not gapped.

Finally, we remark that here we have only considered the standard on-site
hyperfine coupling to the copper spin, and not the transferred hyperfine
coupling which could in principle couple the observed nuclear spin at the dimer
site to the neighboring (strongly polarized) monomer spin. This latter
mechanism typically relies on almost negligible admixture ($\sim$1\,\%) of the
on-site $s$-wave orbital in the extended Wannier wave function belonging to the
neighboring electronic spin. While the on-site spin polarization induced in
this way is negligible ($\sim$1\,\%), very high hyperfine coupling of an
$s$-wave orbital ($\sim$200\,T/$\mu_{\rm{B}}$) can in principle provide
significant transferred hyperfine field. However, this field is positive and
isotropic, in obvious contradiction to what is observed in azurite. Negative
and strongly anisotropic $H_{\rm{spin}}$ necessarily implies significant
on-site spin polarization of the Cu orbital.

To provide a simple discussion for the observed polarization of dimer spins, we
note that the approximate wave functions proposed for the 1/3 plateau of the
diamond chain \cite{Okamoto_JPC03} can be generalized to represent an arbitrary
mixture of the three single-spin-flip states, $\Psi(\mp,\beta) =
\cos\beta\,[$\,$|$$\downarrow \uparrow \uparrow \rangle\,\mp\,$$|$$\uparrow
\downarrow \uparrow \rangle\,]/\sqrt{2} - \sin\beta\,$$|$$\uparrow \uparrow
\downarrow \rangle$, which by construction has correct total and local spin
polarization of $\mu_{\rm{B}}$ and
$\mu_{\rm{B}}\times$($\sin^2$$\beta,~\sin^2$$\beta,~\cos^2$$\beta-\sin^2$$\beta$),
respectively. In particular, there is equal spin polarization on the two dimer
spins. In this notation the two reference plateau states \cite{Okamoto_JPC03}
are $00u$ = $\Psi(-,0)$ and $uud$ = $\Psi(+,\arccos(1/\sqrt{3}))$, where the
sign difference corresponds to the different symmetry. (Note that at least two
unit cells should be taken into account to properly represent all symmetries of
the system.) Using $\Psi(\mp,\beta)$ as a trial function to minimize the energy
by optimizing $\beta$, one can easily see that the pure singlet $00u$ ($\beta =
0$) state is obtained only for the symmetric diamond couplings $J_1 = J_3$,
while deviation from this case necessarily leads to some admixture of the
triplet, meaning some non-zero polarization of the dimer spins ($\beta \neq
0$). Observed polarization on the dimer site means that this admixture is
significant, $\sin\beta \approx \sqrt{0.1} \approx 0.3$ \cite{3rd_state}, and
thus should correspond to an important asymmetry of couplings.  A correct
estimate of the corresponding $J$ coupling values should rely on numerical
solutions of the spin Hamiltonian, relating the observed spin polarization to
the corresponding constraint on the $J$ couplings, say the $J_3/J_1$ vs.
$J_2/J_1$ dependence \cite{private}. Two available points predicting the
correct local spin polarization, $J_3/J_1 = -0.5$ \cite{3rd_state} and +0.4
\cite{private}, show that from NMR we cannot directly conclude if one of the
couplings is ferromagnetic or not. However, if NMR data is combined with other
constrains, as the width of the plateau from the magnetization data
\cite{Kikuchi_PRL05} and the energy of excitations from the neutron scattering
data \cite{Rule_PRL08}, one should clearly define the couplings, or indicate
whether the diamond chain model is too simple to describe azurite. Here we
recall a possible influence of inter chain couplings.

In conclusion, by copper NMR in the 1/3 magnetization plateau of azurite we
have determined the local spin polarization of the dimer spins to be
$\approx0.1\,\mu_{\rm{B}}$. This provides the first direct evidence for a
``quantum'' type of a 1/3 plateau having \textit{no} classical analogue, which
consists of dimers in a singlet state and fully polarized monomers. The
deviation from ideal zero polarization of the dimer implies important asymmetry
of the diamond couplings, $J_1 \neq J_3$, and provides a strong constraint for
the determination of their values.

\begin{acknowledgments}
We acknowledge fruitful discussions with B. Grenier, A. Honecker, O. Janson, F.
Mila, H. Ohta, T. Sakai and T. Tonegawa. Part of this work has been supported
by the French ANR project NEMSICOM and by European Commission through the
Transnational Access Specific Support Action (Contract No. RITA-CT-2003-505474)
and the EuroMagNET network (Contract No. RII3-CT-2004-506239).

\end{acknowledgments}


\begin{references}

\bibitem{Kikuchi_PRL05}
H. Kikuchi \textit{et al.}, Phys. Rev. Lett. \textbf{94}, 227201 (2005).

\bibitem{uud}
T. Ono \textit{et al.}, Phys. Rev. B \textbf{67}, 104431 (2003) and Refs.
therein.

\bibitem{Okamoto_JPC03}
K. Okamoto, T. Tonegawa and M. Kaburagi, J. Phys.: Condens. Matter \textbf{15}
5979 (2003).

\bibitem{Gu_PRB07}
B. Gu and G. Su, Phys. Rev. B \textbf{75}, 174437 (2007).

\bibitem{Comment}
B. Gu and G. Su, Phys. Rev. Lett. \textbf{97}, 089701 (2006); H.~Kikuchi
\textit{et al.}, Phys. Rev. Lett. \textbf{97}, 089702 (2006).

\bibitem{Rule_PRL08}
K. C. Rule \textit{et al.}, Phys. Rev. Lett. \textbf{100}, 117202 (2008).

\bibitem{Mikeska_PRB08}
H.-J. Mikeska and C. Luckmann, Phys. Rev. B \textbf{77}, 054405 (2008).

\bibitem{Kikuchi_PTPS05}
H. Kikuchi \textit{et al.}, Prog. Theor. Phys. Suppl. \textbf{159}, 1 (2005).

\bibitem{Abragam61}
A. Abragam, \emph{The Principles of Nuclear Magnetism}, (Clarendon
Press, Oxford, 1961).

\bibitem{approx}
A minor approximation is to neglect the fact that the strong anisotropy and
values of the ${\sf{A}}$ tensor make ${\bf{H}}_{\rm{eff}}$ and ${\bf{H}}_0$
somewhat non-parallel.

\bibitem{Belokoneva_01}
E. L. Belokoneva, Yu. K. Gubina, and J. B. Forsyth, Phys. Chem. Miner.
\textbf{28}, 498 (2001).

\bibitem{fast_T_2}
Well known example is the relaxation in the superconducting state of high $T_c$
superconductors, for a review see C. Berthier \textit{et al.}, J. Phys. I
France \textbf{6}, 2205 (1996).

\bibitem{3rd_state}
This corresponds closely to the state $\Psi(-,\arcsin(1/3))$ which was
identified at $J_1 : J_2 : J_3 = 1 : 2 : -0.5$ and proposed to be the third
type of plateau state \cite{Gu_PRB07}. However, this state rather appears as
$00u$ state modified by some admixture of the triplets (on dimer spins).

\bibitem{private}
One compatible set of values is $J_1 : J_2 : J_3 = 1 : 1.2 : 0.4$, T. Tonegawa,
private communication.

\end{references}


\end{document}